\documentclass[reprint, superscriptaddress, secnumarabic, amssymb, nobibnotes, aps, prl]{revtex4-1}

\usepackage{graphicx}
\usepackage{epstopdf}
\usepackage[T1]{fontenc}
\usepackage[latin9]{inputenc}
\usepackage{amsbsy}
\usepackage{gensymb}
\usepackage[T1]{fontenc}
\usepackage[latin9]{inputenc}
\usepackage{amsmath}
\usepackage{amssymb}
\usepackage{bbm}
\usepackage{physics}
\usepackage{braket}
\usepackage{xcolor}
\allowdisplaybreaks
\usepackage{graphicx}
\usepackage[colorlinks=true]{hyperref}  
\hypersetup{
    bookmarks=true,         
    unicode=false,          
    pdftoolbar=true,        
    pdfmenubar=true,        
    pdffitwindow=false,     
    pdfstartview={FitH},    
    pdftitle={AgSnSe_2},    
    pdfauthor={},     
    pdfsubject={},   
    pdfcreator={},   
    pdfproducer={}, 
    pdfkeywords={} {} {}, 
    pdfnewwindow=true,      
    colorlinks=true,       
    linkcolor=blue, 
    citecolor=blue,        
    filecolor=magenta,      
    urlcolor=blue           
} 
\usepackage[normalem]{ulem}

\newcommand{\equref}[1]{Eq.~(\ref{#1})}
\newcommand{\equsref}[2]{Eqs.~(\ref{#1}) and (\ref{#2})}

\newcommand{\figref}[1]{Fig.~\ref{#1}}

\renewcommand{\approx}{\simeq}

\begin{document}
\title{\textrm{Superconducting ground state study of valence skip compound AgSnSe$_2$}}

\author{A. Kataria}
\affiliation{Department of Physics, Indian Institute of Science Education and Research Bhopal, Bhopal, 462066, India}
\author{Arushi}
\affiliation{Department of Physics, Indian Institute of Science Education and Research Bhopal, Bhopal, 462066, India}
\author{S.~Sharma}
\affiliation{Department of Physics and Astronomy, McMaster University, Hamilton, Ontario L8S 4M1, Canada}
\author{T. Agarwal}
\affiliation{Department of Physics, Indian Institute of Science Education and Research Bhopal, Bhopal, 462066, India}
\author{M.~Pula}
\affiliation{Department of Physics and Astronomy, McMaster University, Hamilton, Ontario L8S 4M1, Canada}
\author{J.~Beare}
\affiliation{Department of Physics and Astronomy, McMaster University, Hamilton, Ontario L8S 4M1, Canada}
\author{S. Yoon}
\affiliation{TRIUMF, Vancouver, British Columbia V6T 2A3, Canada}
\affiliation{Department of Physics, Sungkyunkwan University, Suwon 16419, Korea}
\author{Y. Cai}
\affiliation{TRIUMF, Vancouver, British Columbia V6T 2A3, Canada}
\affiliation{Quantum Matter Institute, The University of British Columbia, Vancouver, BC V6T 1Z4, Canada}
\author{K. M. Kojima}
\affiliation{TRIUMF, Vancouver, British Columbia V6T 2A3, Canada}
\author{G.~M.~Luke}
\affiliation{Department of Physics and Astronomy, McMaster University, Hamilton, Ontario L8S 4M1, Canada}
\affiliation{TRIUMF, Vancouver, British Columbia V6T 2A3, Canada}	
\author{R.~P.~Singh}
\email[]{rpsingh@iiserb.ac.in} 
\affiliation{Department of Physics, Indian Institute of Science Education and Research Bhopal, Bhopal, 462066, India}

\begin{abstract}
The valence-skipped superconductors are natural candidates for unconventional superconductivity, as they can exhibit a negative effective, attractive interaction for electron-pairing. This work reports comprehensive XRD, magnetization, specific heat and muon spin rotation and relaxation measurements ($\mu$SR) on a valence-skipped compound: AgSnSe$_2$. The temperature dependence of the electronic specific heat ($C_{el}(T)$) and  of the upper critical field ($H_{c2}(T)$) provide evidence of two-gap superconductivity, which is also confirmed by our transverse-field $\mu$SR measurements. Our zero-field $\mu$SR measurements suggest preserved time-reversal symmetry in the superconducting ground state of AgSnSe$_2$.

\end{abstract}
\maketitle
\section{Introduction}

Understanding the microscopic pairing mechanism of unconventional superconductors is one of the most challenging problems in condensed matter physics. It has recently attracted much interest due to its possible applications in quantum technologies. In unconventional superconductors, the electron-pair formation is not mediated by phonons but by different mechanisms, which may include charge, spin or magnetic fluctuations \cite{sf,mf,cf}.

Valence-skipped materials are an exciting class of materials which may become unconventional superconductors. In these materials, due to valence-skipping/valence fluctuations, negative-$U$ centers are formed, which causes an attractive on-site interaction between the electrons \cite{hf,nsc,cm,as}, and promotes superconductivity with relatively high $T_c$ compared to conventional superconductors \cite{htc,htc2,vs,vs2,vs3,vs4}. This valence-skipped induced negative-$U$ superconductivity has previously been reported in Tl-doped PbTe, K- and Na-doped BaBiO$_3$ \cite{tlpt,BBO3,kbbo3}, with a high $T_c$ value. Valence-skipped materials also exhibit intriguing unconventional properties and other quantum phenomena, including charge density wave (CDW), pseudogap, and charge Kondo effect \cite{ke,hf,cf}. However, the exact superconducting pairing mechanism in these valence-skipped compounds is still elusive. With the possibility of achieving a high value of $T_c$ in relatively low carrier density materials, a valance-skipped superconducting pairing mechanism could be a novel root to realize high $T_c$ materials, motivating detailed microscopic studies which are not available for these materials to date.

The metal chalcogenide superconductor, AgSnSe$_2$, where Ag partially substitutes Sn in SnSe, is a novel valence-skipped system \cite{ass2,ass2sc}. The $ab$-$initio$ band structure calculations indicate SnSe is a  nontrivial crystalline topological material \cite{ss1,ss}. X-ray absorption (XAS) and x-ray photoemission spectroscopy (XPS) studies of AgSnSe$_2$ suggested the presence of valency states Sn$^{2+}$ and Sn$^{4+}$, which might create dynamic negative-$U$ centers and contribute to superconductivity \cite{xps}. Further, $^{119}$S\"{n} M\"{o}ssbauer spectroscopy suggests valence skip behaviour with excess Sn \cite{moss}. The coexisting superconductivity and nontrivial band topology in AgSnSe$_2$ make it a unique valence-skipped system and motivate further studies of this compound. A thorough microscopic investigation requires understanding the correlation between unconventional superconductivity, valence fluctuations and the effect of nontrivial band topology on the superconducting ground state of AgSnSe$_2$ and valence-skipped compounds in general.

In this paper, we report the bulk and microscopic superconducting properties of AgSnSe$_2$ via magnetization, specific heat and muon spin rotation and relaxation ($\mu$SR) measurements. Our measurements confirm the superconducting transition temperature $T_c$ = 4.91(2) K. Transverse-field (TF) $\mu$SR measurements, along with the temperature dependence of the specific heat and the upper critical field suggest a two isotropic $(s + s)$ superconducting gap structure. Zero-field (ZF) muon spin relaxation measurements indicate that within our resolution, time-reversal symmetry (TRS) is preserved in the superconducting state of AgSnSe$_2$.

\section{Experimental Details}

The polycrystalline sample was synthesized using the solid-state reaction method. The required elemental powders Ag (99.999\%), Sn (99.99\%), and Se (99.999\%) were mixed in a stoichiometric ratio. The mixture was then palletized and heated at 800$\degree$C for 48 hours in a vacuum-sealed tube, followed by water quenching. At room temperature, the powder x-ray diffraction (XRD) pattern was collected using a PANalytical diffractometer equipped with Cu$K_{\alpha}$ radiation ($\lambda$ = 1.5406 \AA). Magnetic measurements were performed using a Quantum Design MPMS XL SQUID magnetometer with RSO (reciprocating sample option) insert, and for measurements down to 0.5 K, an IQuantum He$^{3}$ insert was used. A physical property measurement system (PPMS, Quantum Design) was used to measure the specific heat of the sample. Zero-field and transverse-field $\mu$SR measurements were performed at TRIUMF, Center for Molecular and Materials Science, Vancouver, Canada \cite{msr}. In the TF geometry, the applied magnetic field was perpendicular to the initial muon spin direction. The $\mu$SR data were analyzed using the musrfit software \cite{musrfit}.  

\section{Results and Discussion}
\begin{figure}
\centering
\includegraphics[width=0.92\columnwidth, origin=b]{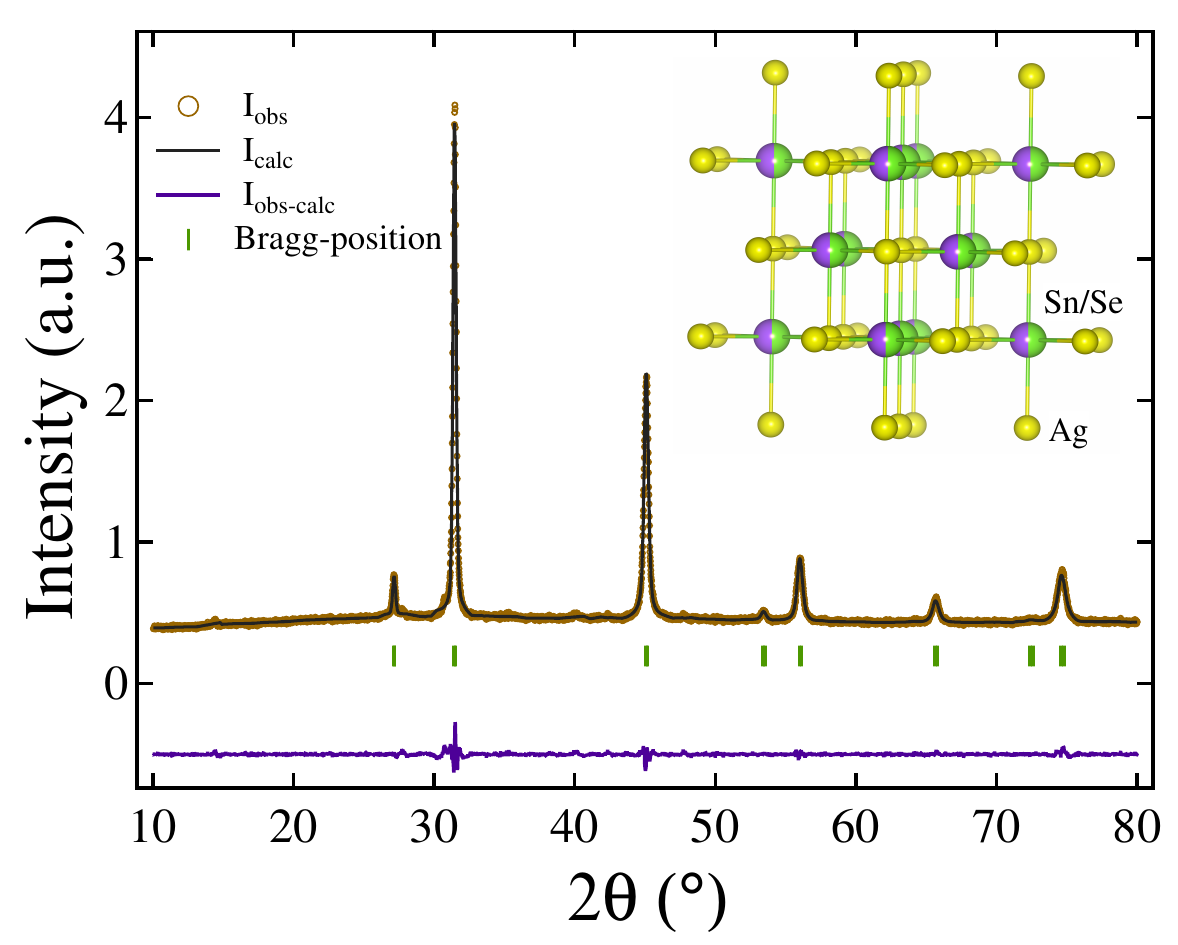}
\caption{\label{Fig1:char} Refined room temperature powder XRD pattern of AgSnSe$_2$ with inset showing the NaCl crystal structure with equal probability of occupying edge state by Ag and Sn atoms.}
\end{figure}
The crystal structure of AgSnSe$_2$ is shown in the inset of \figref{Fig1:char} depicting the partial substitution of Ag at the Sn site in SnSe. Rietveld refined powder XRD pattern of AgSnSe$_2$ is shown in \figref{Fig1:char}, confirming the crystallization in a cubic NaCl structure with $Fm$-$3m$ space group. The obtained lattice parameters from the refinement, a = b = c = 5.6865(2) $\text\AA$, are in good agreement with previously reported values \cite{str}.

\subsection{Magnetization}

\begin{figure*}[ht!]
\centering
\includegraphics[width=2.02\columnwidth, origin=b]{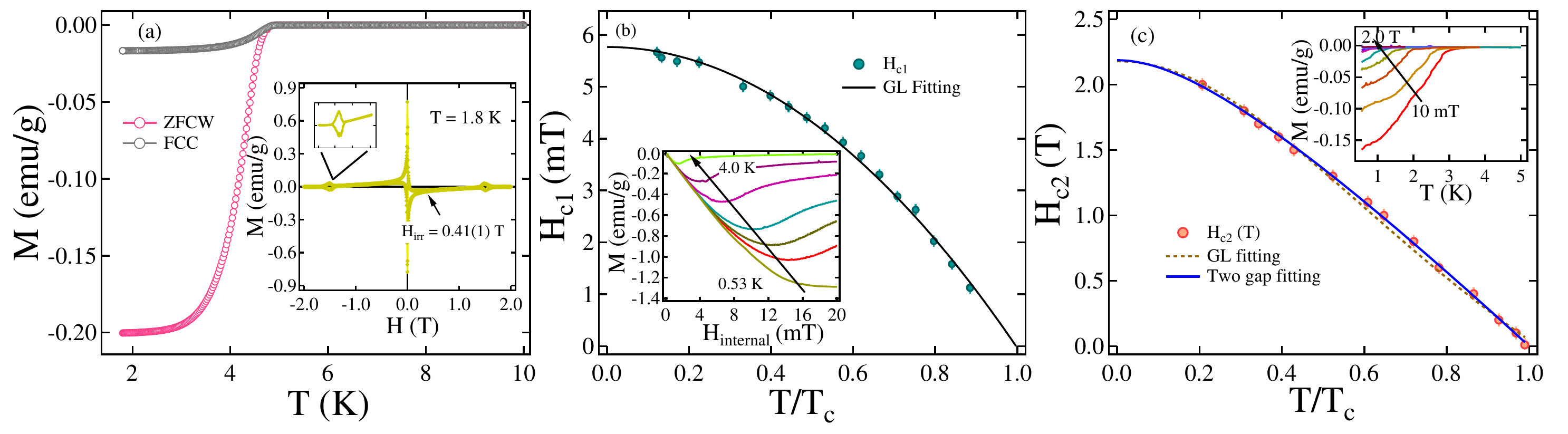}
\caption{\label{Fig2:mt} (a) The magnetic measurements under ZFCW and FCC mode with the magnetic hysteresis loop at 1.8 K in the inset. (b) The temperature variation of the lower critical field and the inset shows the $M$ dependence on the internal magnetic field. (c) The upper critical field temperature variation fitted by the GL equation and the two-gap model are represented by dashed and solid lines, respectively. Inset shows the temperature-dependent magnetization at the different applied fields.}
\end{figure*}

The temperature-dependent magnetic measurements are performed in zero-field-cooled warming (ZFCW) and field-cooled cooling (FCC) mode under 1 mT magnetic field, as shown in \figref{Fig2:mt}(a). The observed superconducting transition temperature of AgSnSe$_2$, $T_c$ = 4.91(2) K, is close to previously reported values \cite{ass2,ass2sc}. The separation between ZFCW and FCC curves indicates the presence of strong flux pinning. The magnetization variation under a high magnetic field at temperature 1.8 K (inset of \figref{Fig2:mt}(a)) also suggests strong pinning with complex vortex nature as a small area under the magnetization loop with the fish-tail effect is observed \cite{fte}. An irreversible magnetic field is observed at $H_{irr}$ = 0.41(1) T, above which vortices start to unpin. The field-dependent magnetization measurements at different temperatures below $T_c$ are used to estimate the lower critical field value. 

We performed the magnetization measurements on a spherical sample piece, allowing accurate demagnetising field determination. We calculate the internal field of the sample by using $H_{internal}$ = $H_{applied}$ - $N M$, where $N$ = 1/3 is the demagnetising factor of a sphere, and $M$ is the magnetization \cite{Coh_Leng}. The inset of \figref{Fig2:mt}(b) shows the variation of $M$ versus $H_{internal}$, and the point of linear deviation in the low-field region is considered as $H_{c1}$ for the respective isotherm. The temperature dependence of the lower critical field, $H_{c1}$, is well described using the Ginzburg-Landau (GL) equation,
\begin{equation}
H_{c1}(T) = H_{c1}(0)\left[1-\left(\frac{T}{T_c}\right)^2\right]
\label{eqn1:HC1}
\end{equation} 
where the lower critical field is estimated to be $H_{c1}(0)$ = 5.76(6) mT (\figref{Fig2:mt}(b)). 

From the magnetization measurement, the shift in $T_c$ to a lower value with an increasing magnetic field is noted to extract the upper critical field (inset of \figref{Fig2:mt}(c)). We attempted to fit the $H_{c2}$ versus $T$ curve by the GL equation,
\begin{equation}
H_{c2}(T) = H_{c2}(0)\left[\frac{(1-t^2)}{(1+t^2)}\right]
\label{eqn2:HC2}
\end{equation} 
where $t = T/T_c$ is the reduced temperature. \figref{Fig2:mt}(c) shows the relatively poor fitting of the $H_{c2}$ curve from the single-band GL model providing $H_{c2}^{GL}$(0) = 2.13(3) T. Therefore, a two-band model was used to analyze the $H_{c2}$ variation. The two-band equation for the upper critical field is written in the parametric form as \cite{mb2,pts2,txs,lfo,ns2},
\begin{multline}
\ln t= -\frac{1}{2}\left[U(s)+U(\eta s)+\frac{\lambda_0}{w}\right] +
\\
\left(\frac{1}{4}\left[U(s)-U(\eta s)-\frac{\lambda_-}{w}\right]^2+\frac{\lambda_{eh}\lambda_{he}}{w^2}\right)^{1/2},
\label{eqn3:HC2}   
\end{multline}
where
\begin{equation}
U(s)=\psi\left(s+\frac{1}{2}\right)-\psi\left(\frac{1}{2}\right),
\label{eqn4:psi}
\end{equation}
$H_{c2} = \frac{2 \Phi_0 T_c th}{D_e}$, $\eta = \frac{D_h}{D_e}$ and the parameter $h$ changes from 0 to 1 as $T$ varies from $T_c$ to 0 with $t$ being the reduced temperature. The variables  $\lambda_{ee},\lambda_{hh},\lambda_{he},\lambda_{eh}$ are the matrix elements of the BCS coupling constants and $\lambda_{-} = \lambda_{ee}-\lambda_{eh}$, $\lambda_0 = \lambda_{-}^2 +\lambda_{eh}\lambda_{he}$, $w = \lambda_{ee}\lambda_{hh} -\lambda_{he}\lambda_{eh}$.  $D_e$ and $D_h$ are the electron and hole diffusivity, respectively while $\Phi_0$ is the magnetic flux quantum and $\psi$ is the digamma function. The fit of the $H_{c2}$ curve using the two-band theory yields $H_{c2}^{2G}$(0) = 2.18(7) T. The obtained values of the upper critical field from the two theories are much smaller than the Pauli paramagnetic limit, expressed as $H_{c2}^P(0)= 1.86 T_c = 9.13(3)$T \cite{Pauli_1,Pauli_2}. Further, Ren $\textit{et al.}$ \cite{ass2sc} reported the anomalous broadening of the magnetic-field-induced resistive transition and an increase in the upper critical field value in the low-temperature region in AgSnSe$_2$ which might be associated with multiple superconducting gap presence, as suggested by our upper critical field results and also observed in MgB$_2$ \cite{mgb22}.

We can evaluate various superconducting parameters using our $H_{c2}$ and $H_{c1}$ values. The superconducting coherence length, $\xi_{GL}$(0) is calculated using GL theory as \cite{Coh_Leng}, $H_{c2}(0)=\left(\frac{\Phi_0}{2\pi\xi_{GL}(0)^2}\right) $, providing $\xi_{GL}$(0) = 12.2(4)~nm. The London penetration depth, $\lambda_{GL}$ is obtained from the lower critical field, $H_{c1}$(0) using the expression \cite{lam},
\begin{equation}
H_{c1}(0) = \frac{\Phi_{0}}{4\pi\lambda_{GL}^2(0)}\left(\mathrm{ln}\frac{\lambda_{GL}(0)}{\xi_{GL}(0)}+0.12\right)   
\label{eqn5:PD}
\end{equation} 
which gives $\lambda_{GL}$(0) = 309(13) nm for $H_{c1}$(0) = 2.07(1) mT and $\xi_{GL}$(0) = 12.2(4)~nm. The GL parameter, $\kappa_{GL} = \lambda_{GL}(0)/\xi_{GL}(0)$ = 25(2), is indicating the strong type-II superconducting nature in AgSnSe$_2$. Using these parameters and the relation $H_{c1}H_{c2} = H_{c}^2\mathrm{ln}\kappa_{GL}$ \cite{lam}, the thermodynamic critical field value, $H_c(0)$ = 62(7) mT is also estimated.

\subsection{Specific heat}

\begin{figure} 
\includegraphics[width = 1.00\columnwidth, origin=b]{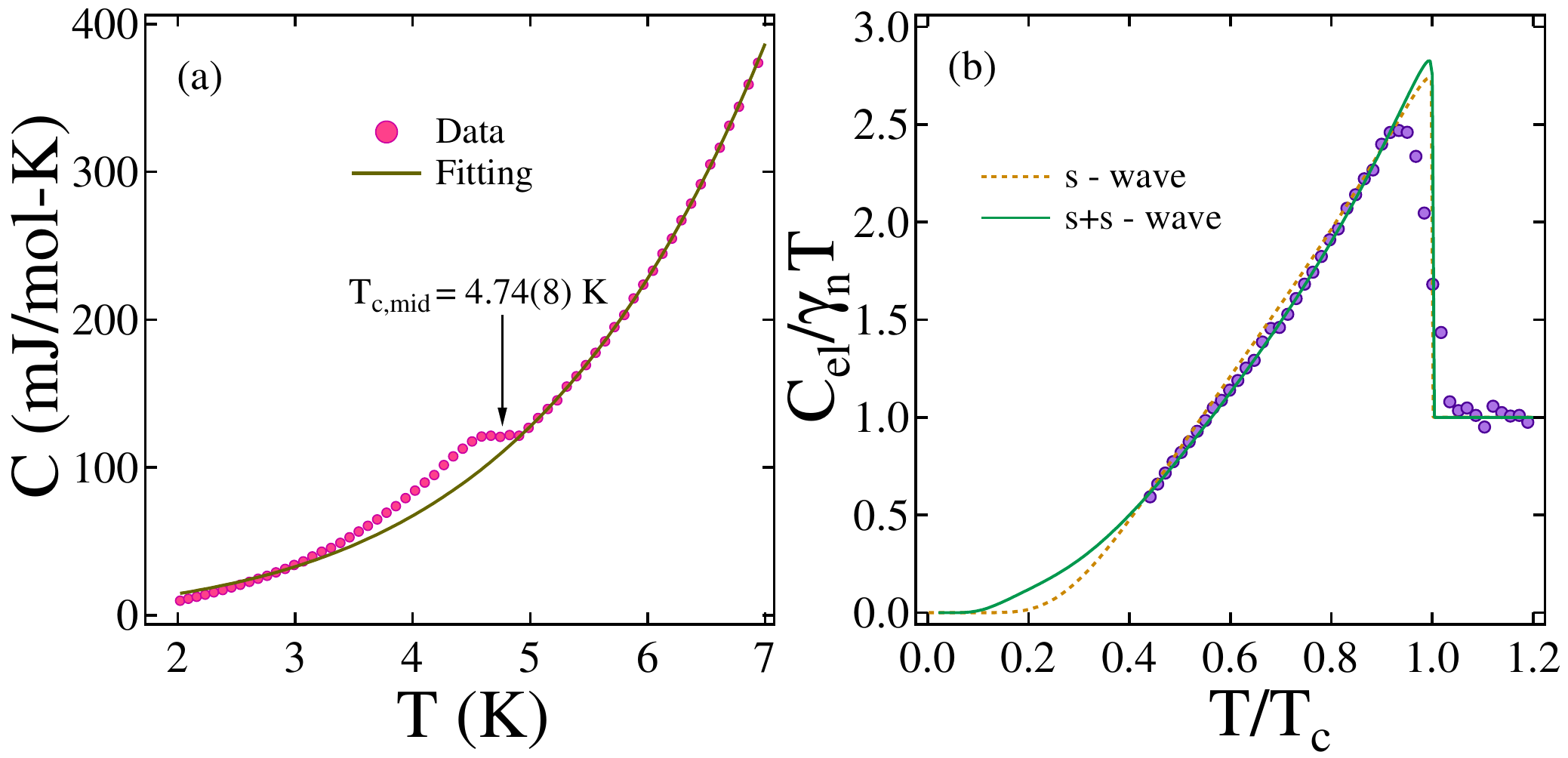}
\caption{ \label{Fig3:SH} (a) Specific heat fitting in the normal region. (b) Normalized electronic specific heat data, $C_{el}/\gamma_{n}T$, fitted using two different models $s$-wave shown by the dotted yellow line and two-gap $s + s$-wave shown by the solid green line.}
\end{figure}

The temperature-dependent specific heat of AgSnSe$_2$ under zero magnetic field is shown in the \figref{Fig3:SH}(a). The significant jump in the specific heat around temperature $T_{c,midpoint}$ = 4.74(8) K further confirms the bulk nature of superconductivity in our sample. The normal-state specific heat above $T_c$ was fit by the Debye relation,
\begin{equation}
C=\gamma_n T +\beta_3 T^3 +\beta_5 T^5
\label{eq6:SH}
\end{equation}
where $\gamma_n T$ is the electronic contribution, $\beta_3T^3$ is the phononic contribution and $\beta_5T^5$ is the anharmonic contribution in the specific heat. The best fit to the data yields Sommerfeld coefficient, $\gamma_n$ = 4.6(1) mJ$\cdot$mol$^{-1}$ K$^{-2}$, Debye constant, $\beta_3$ = 0.68(1) mJ$\cdot$mol$^{-1}$ K$^{-4}$ and $\beta_5$ = 7.5(1) $\mu$J$\cdot$mol$^{-1}$ K$^{-5}$. 

Further, $\gamma_n$  is used to evaluate the density of states at the Fermi level, $D_{C}(E_{\mathrm{F}})$ via the relation, $\gamma_{n}= \left(\frac{\pi^{2}k_{B}^{2}}{3}\right)D_{C}(E_{\mathrm{F}})$, where $k_{B}$ = 1.38$\times$10$^{-23}$ J K$^{-1}$. $D_{C}(E_{\mathrm{F}})$ is calculated to be 1.97(6) states eV$^{-1}$f.u$^{-1}$. The Debye temperature is obtained from $\beta_3$ using the expression, $\theta_{D}= \left(\frac{12\pi^{4}RN}{5\beta_{3}}\right)^{\frac{1}{3}}$ = 255(3) K, where $R$ = 8.314 J mol$ ^{-1} $K$ ^{-1} $ is the gas constant. From $\theta_D$ and $T_c$, the electron-phonon coupling constant, $\lambda_{e-ph}$ can also be calculated from the inverted McMillan's equation \cite{McMillan}, 
\begin{equation}
\lambda_{e-ph} = \frac{1.04+\mu^{*}\mathrm{ln}(\theta_{D}/1.45T_{c})}{(1-0.62\mu^{*})\mathrm{ln}(\theta_{D}/1.45T_{c})-1.04 }
\label{eqn7:Lambda}
\end{equation}
here $\mu^{*}$ is the screened Coulomb repulsion and is taken as 0.13. For $\theta_{D}$ = 255 K and $T_{c,midpoint}$ = 4.74 K, resulting $\lambda_{e-ph} $ = 0.69(1). The obtained value of $\lambda_{e-ph}$ for AgSnSe$_2$ is in the range of moderately coupled superconductors.

To evaluate the superconducting gap parameter, we analyzed the electronic-specific heat below $T_c$. The electronic contribution is extracted by subtracting the lattice contribution from the total specific heat. The normalized superconducting gap, $\Delta(0)/k_B T_c$ is estimated by fitting the data to the BCS expectation of electronic-specific heat, which is expressed as \cite{cel},
\begin{equation}
\frac{C_{el}}{\gamma_n T_{c}} = t\frac{d}{dt}\left[\frac{-6\Delta(0)}{\pi^2k_{B}T_{c}}\int_{0}^{\infty}[ \textit{f}\ln(f)+(1-f)\ln(1-f)]
\\ dy \right]\\
\label{eqn8:BCS1}
\end{equation}
\\
here $ t = T/T_c$, $\textit{f}$($\xi$) = [exp($\textit{E}$($\xi$)/$k_{B}T$)+1]$^{-1}$ is the Fermi function, $\textit{E}$($\xi$) = $\sqrt{\xi^{2}+\Delta^{2}(t)}$, with $E(\xi$) being the energy of the normal electrons measured relative to the Fermi energy. $\textit{y}$ = $\xi/\Delta(0)$ and $\Delta(t)$ = tanh$\{1.82(1.018[(\mathit{1/t}) - 1])\}^{0.51}$ is the BCS approximation for the temperature dependence of energy gap. The normalized $C_{el}(T)$ fitting deviates from a single $s$-wave model, as shown in the \figref{Fig3:SH}(b), whereas a two-gap isotropic $s + s$ model fits the data better. The same two-gap model has also been used in other multi-gap superconductors such as MgB$_2$ \cite{mgb2}, Lu$_2$Fe$_3$Si$_5$ \cite{l2f3s5}, and La$_7$Ni$_3$ \cite{l7n3}. In this model, the two parameters $\Delta_1(0)/k_B T_c$ and $\Delta_2(0)/k_B T_c$ correspond to the two different superconducting gaps, and the weighted contribution, $x$ of the partial Sommerfeld coefficient are the three fitting parameters involved. Best fitting of the data yields $x$ = 0.17(6), $\Delta_1(0)/k_B T_c$ = 0.80(8) and $\Delta_2(0)/k_B T_c$ = 2.16(3) (\figref{Fig3:SH}(b)). A significant difference in the two gap values and gap ratio $\frac{\Delta_2(0)}{\Delta_1(0)}$ $\approx$ 2.7(3) is observed with the smaller gap, $\Delta_1$ having relatively weak contribution (0.17) in the superconducting state. The larger gap value, $\Delta_2(0)/k_B T_c$ = 2.16, is higher than the BCS value in the weak coupling limit, suggesting  strong electron-phonon coupling in AgSnSe$_2$. A previous report on polycrystalline AgSnSe$_2$ reported a single isotropic gap (also somewhat larger than that expected in the weak coupling limit of BCS theory); it is possible that the apparent absence of a two-gap nature in that study might be due to sample differences or could indicate that a multi-gap analysis was not performed \cite{ass2sc}.

\subsection{Muon spin rotation and relaxation}
 
\begin{figure}[b]
\includegraphics[width=1.00\columnwidth]{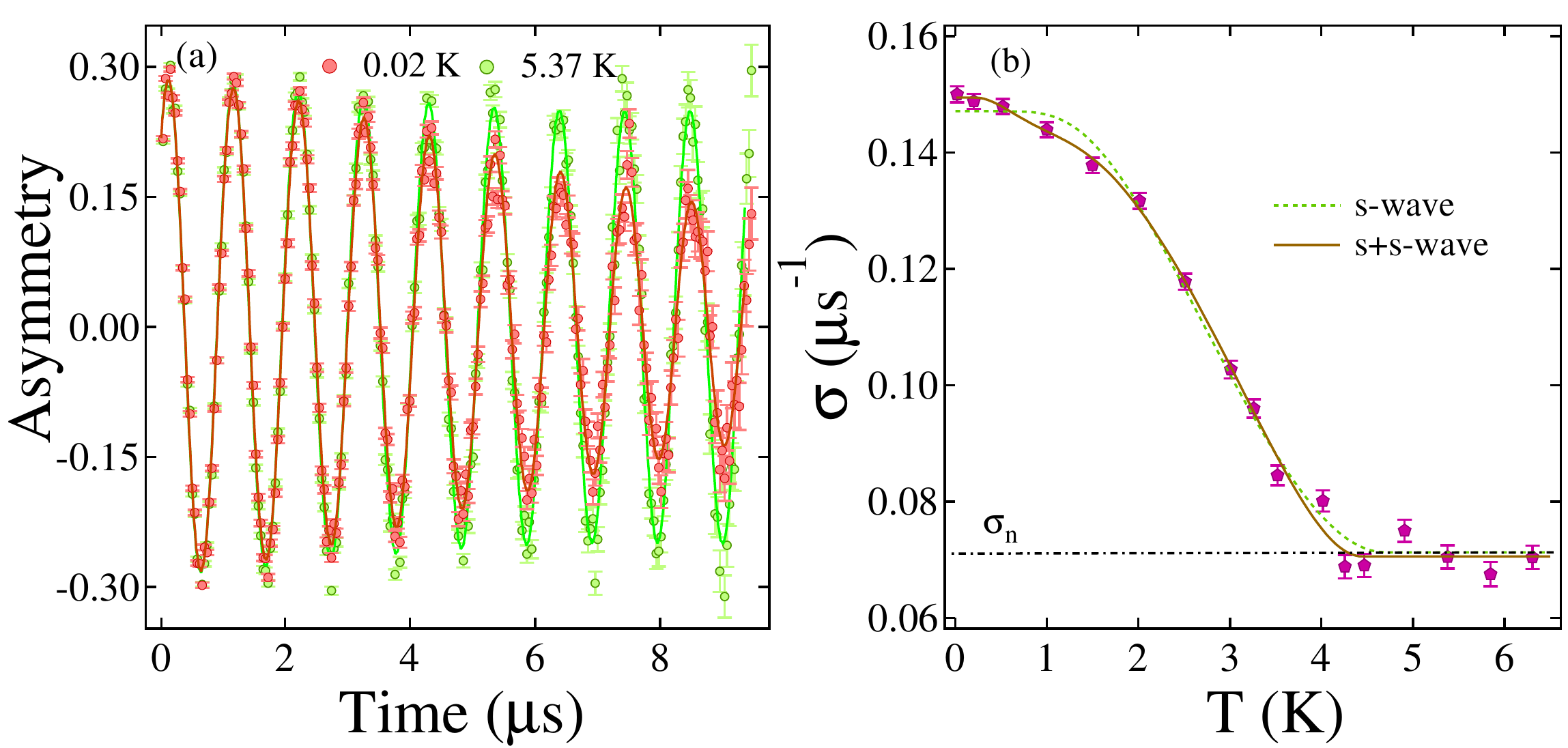}
\caption{\label{Fig4:TF} (a) Transverse-field spectra recorded in an applied magnetic field of 0.1 T at temperature 5.37 K ( > $T_{c}$)  and 0.02 K (< $T_{c}$) plotted in a rotating reference frame of 0.095 T, where solid lines are the fits using \equref{eq9:tf}. (b) Temperature dependence of total relaxation rate, $\sigma$. The dashed line represents the $s$-wave fitting, the solid line represents the two-isotropic $s + s$-wave model, and $\sigma_n$ is the fitting parameter.}
\end{figure}

To further understand the superconducting gap structure of the valence-skipped superconductor AgSnSe$_2$ at the microscopic level, we carried out transverse-field $\mu$SR measurements. The measurements were performed in the field-cooled protocol in a field of 0.1~T. Representative TF asymmetry spectra above and below  $T_c$ are plotted in a rotating reference frame of 0.095~T for clarity as shown in \figref{Fig4:TF}(a), where a clear difference between the two is observable. Below $T_c$, the precession signal shows significant damping with time due to the  inhomogeneous field distribution from flux-line-lattice (FLL) in the mixed state. The time domain spectra were fit to analyze the temperature dependence of superconducting parameters. The TF asymmetry spectra are described by two sinusoidally oscillating functions, with the Gaussian decaying component characterizing the sample signal and a temperature-independent weakly decaying exponential component characterizing the background signal,
\begin{multline}
A_{TF}(T) = A [F \mathrm{exp}\left(\frac{-\sigma^2 t^2}{2}\right)\mathrm{cos}(wt+\phi)  +
\\(1-F) \mathrm{exp}(-\sigma_{bg})\mathrm{cos}(w_{bg}t+\phi)] 
\label{eq9:tf}
\end{multline}
where $\phi$ is the initial phase of the muons entering the sample and $A$ is the total TF asymmetry. $F$ is the fraction of the signal coming from the sample, and $w,w_{bg}$ are the frequencies of the oscillatory muon signal in the sample and background, respectively. $\sigma$ is the total Gaussian muon relaxation rate. Above the transition temperature $T_c$, $\sigma$ becomes temperature independent, reflecting the nuclear contribution in the normal state called $\sigma_n$ (represented by a horizontal dashed line in \figref{Fig4:TF}(b)). The total relaxation rate, $\sigma$ consists contribution of both superconducting and nuclear parts and can be written as, 
\begin{equation}
 \sigma^{2} = \sigma_{sc}^2+\sigma_{n}^2\label{eq10:sigma},  
\end{equation}
 where $\sigma_{sc}$ is the relaxation in the vortex state. 

The temperature-dependence of the superconducting muon depolarization rate, $\sigma_{sc}$, can be expressed in the semiclassical approximation as,
\begin{multline}
\frac{\sigma_{sc}(T,\Delta_{0,i})}{\sigma_{sc}(0,\Delta_{0,i})}  =
\frac{\lambda_{sc}^{-2}(T,\Delta_{0,i})}{\lambda_{sc}^{-2}(0,\Delta_{0,i})}  =
\\
1 +\frac{1}{\pi}\int_0^{2\pi} \int_{\Delta(T,\phi)}^\infty \left(\frac{\delta f}{\delta E}\right)\frac {EdEd\phi}{\sqrt{E^2-\Delta_i(T,\phi)^2}}
\label{eq12:s}
\end{multline}
where $f(E) = [1+ \exp(E/k_BT)]^{-1}$ is the Fermi function and $\Delta_{i}(T,\phi)= \Delta_{0,i}\delta(T/T_c)g(\phi)$. The temperature variation $\delta(T/T_c) = \tanh [1.82[1.018(T_c/T-1)]^{0.51}]$ and $g(\phi)$ refers to the angular dependence of the superconducting gap function having azimuthal angle $\phi$. 

For an isotropic $s$-wave gap, $g(\phi)$ becomes equal to one. Combining \equsref{eq10:sigma}{eq12:s}, the $\sigma$ variation with temperature can be fitted with $\sigma_n$ as a fitting parameter. The $\sigma$ versus T curve fitted with various models is shown in the \figref{Fig4:TF}(b). The isotropic $s$-wave model with a gap value $\Delta_0$ = 0.68 (5) meV and $\chi^2$ = 1.43 does not properly fit the temperature variation of $\sigma$. Hence a function consisting weighted linear combination of two distinct superconducting gaps is considered, which corresponds to the model used to fit the specific heat data \cite{l2f3s5,l7n3},
\begin {equation}
\frac{\sigma_{sc}(T)}{\sigma_{sc}(0)} = x \frac{\sigma_{sc}(T,\Delta_{0,1})}{\sigma_{sc}(0,\Delta_{0,1})} + (1-x) \frac{\sigma_{sc}(T,\Delta_{0,2})}{\sigma_{sc}(0,\Delta_{0,2})}\\   
\label{eq13:ts}
\end{equation}
where $\Delta_{0,1}$ and $\Delta_{0,2}$ are the two gap values, respectively, and $x$ is the weighting factor, measuring the relative contributions to superconducting parameters. The two-gap $s+s$ wave model fits well than the single gap $s$-wave, for $x$ = 0.20(4), with the respective gap values of $\Delta_{0,1}(0)$ = 0.29(6) meV ($\Delta_{0,1}(0)/k_B T_c$ = 0.77(9) and $\Delta_{0,2}(0)$ = 0.80(5) meV ($\Delta_{0,2}(0)/k_B T_c$ = 2.12(5)) and $\chi^2$ = 1.25 (\figref{Fig4:TF}(b)). The ratio of superconducting gaps $\frac{\Delta_{0,2}(0)}{\Delta_{0,1}(0)}$ $\approx$ 2.7(5) is consistent with the value estimated from the specific heat measurement. 

For an ideal vortex lattice, the relation between the magnetic penetration depth, $\lambda$ and the superconducting depolarization rate $\sigma_{sc}$ for GL parameter $\kappa$ $\geq$ 5 is given by \cite{SigSC_2,sigmsc_3},
\begin {equation}
\frac{\sigma_{sc}^2 (T)}{\gamma_{\mu}^2} =  \frac{0.00371\Phi_0^2}{\lambda^4(T)}
\label{eq11:lam}
\end{equation}
where $\gamma_{\mu}/2\pi$ = 135.5 MHz/T is the muon gyromagnetic ratio and $\Phi_0$ = 2.068 $\times$ 10$^{-15}$ Wb is the magnetic flux quantum. The estimated magnetic penetration depth at $T$ = 0 K is $\lambda(0)$ = 901(35)~nm. The obtained value is significantly different from the value estimated from magnetization, probably due to the different approximation considerations while calculating. Such a large difference in $\lambda$ has also been observed in some other compounds \cite{RRP,Z2I}.

\begin{figure}
\includegraphics[width=0.92\columnwidth]{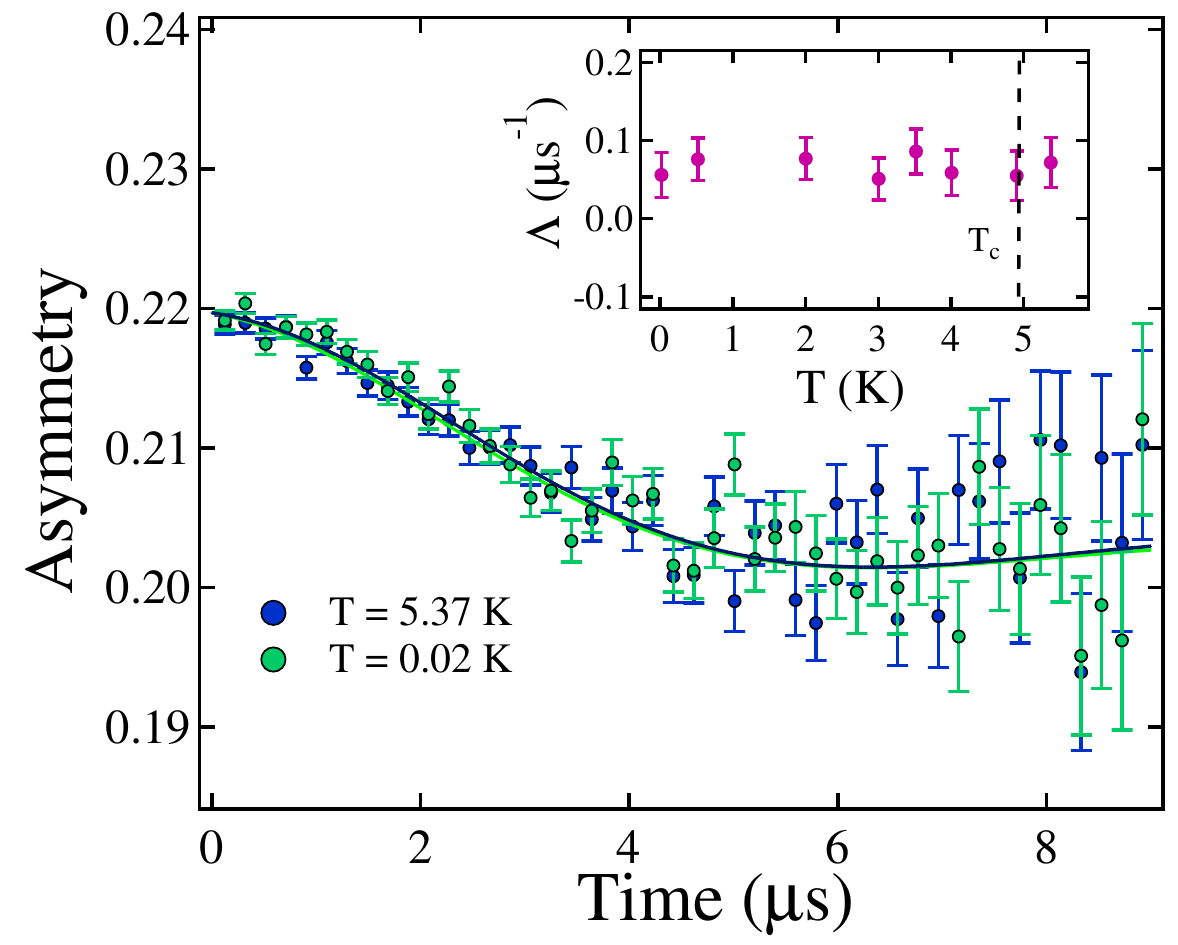}
\caption{\label{Fig5:ZF} Zero-field asymmetry spectra collected below ($T$ = 0.02 K) and above ($T$ = 5.37 K) the transition temperature, $T_{c}$ where the solid lines represent the fit to the data using \equsref{eqn14:ZF2}{eqn15:ZF}. The inset shows the temperature variation of the electronic relaxation rate, $\Lambda$.}
\end{figure}

To further investigate the superconducting ground state of AgSnSe$_2$,  ZF-$\mu$SR measurements were also performed. The ZF-$\mu$SR time domain spectra were measured at various temperatures on either side of the superconducting transition temperature. \figref{Fig5:ZF} shows representative asymmetry spectra above and below $T_c$. The ZF-$\mu$SR spectra were analyzed by fitting the time-domain asymmetry variation using the damped Gaussian Kubo-Toyabe function ($G_{\mathrm{KT}}$) \cite{KT},
\begin{multline}
A(t) = A_{0}[f G_{\mathrm{KT}}(t) \mathrm{exp}(-\Lambda t) + \\(1-f) \mathrm{exp}(-\lambda_{BG}t)]  +  A_{BG},
\label{eqn14:ZF2}   
\end{multline}
where
\begin{equation}
G_{\mathrm{KT}}(t) = \frac{1}{3}+\frac{2}{3}(1-\Delta^{2}t^{2})\mathrm{exp}\left(\frac{-\Delta^{2}t^{2}}{2}\right)
\label{eqn15:ZF}
\end{equation}
and $A_0$ and $A_{BG}$ are the initial asymmetries corresponding to the sample and the background, respectively. $\Delta$ represents the muon spin relaxation due to the randomly oriented, static nuclear moments experienced at the muon site and $\Lambda$ accounts for the electronic relaxation rate. The temperature independence of $\Lambda$ is shown in the inset of \figref{Fig5:ZF}, confirming the absence of any spontaneous internal magnetic field in the superconducting state within our detection limit. The RMS deviation in value of $\Lambda$ provides an upper limit on a spontaneous magnetic field of 0.01 mT. The obtained value is smaller than the value reported for many TRS-breaking superconductors \cite{s2ro4}. However, there is still the possibility of time-reversal symmetry breaking with a spontaneous field less than 0.01 mT; such a small TRS breaking internal field has been previously observed in some superconductors such as UPt$_3$, Re$_6$Hf, La$_7$Ni$_3$ \cite{upt3,re6hf,l7n3}.   

\begin{figure}
\includegraphics[width=0.92\columnwidth]{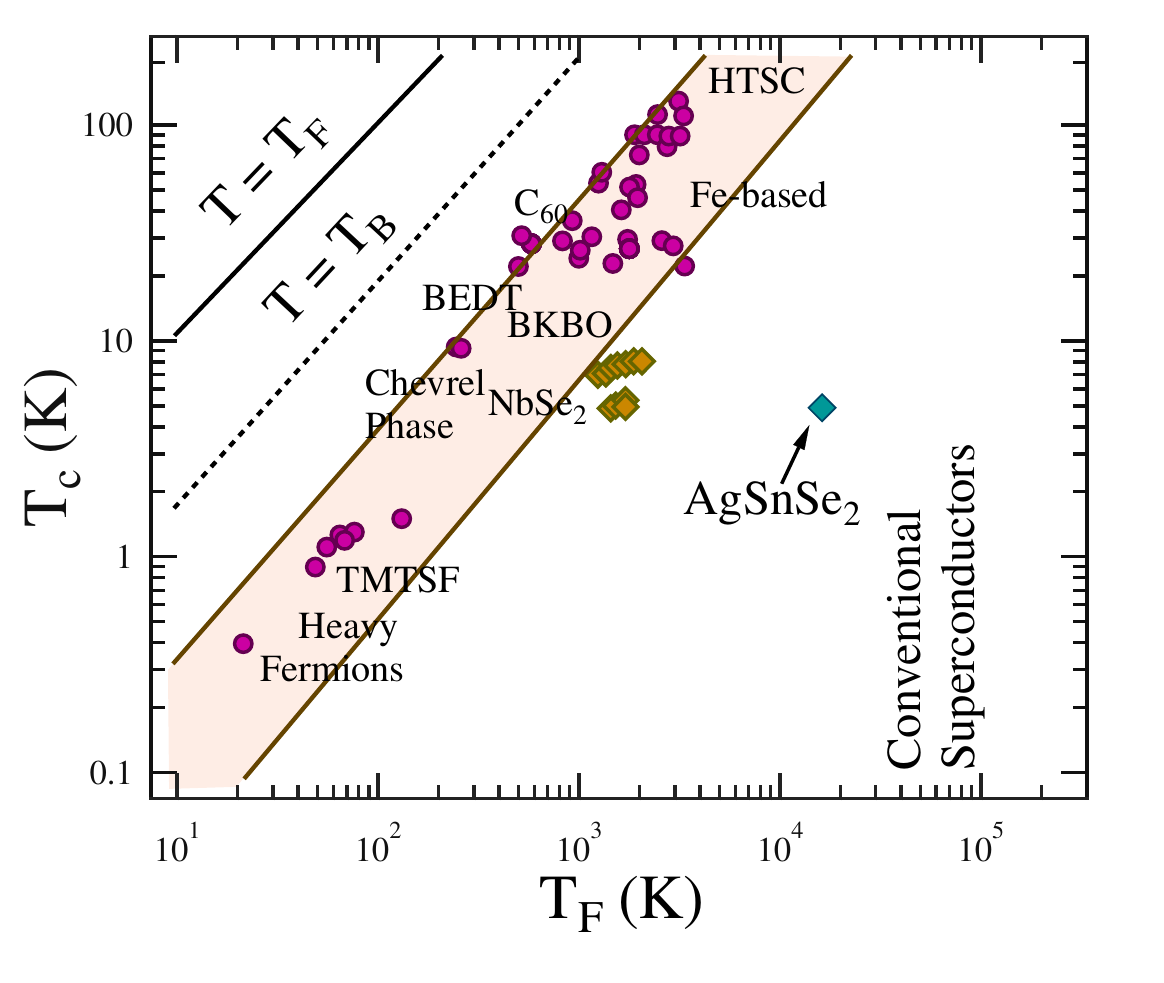}
\caption{\label{Fig6:UP} A plot between the superconducting transition temperature $T_{c}$ and the effective Fermi temperature $T_{\mathrm{F}}$ with AgSnSe$_2$. The data points between the shaded region represent the band of unconventionality \cite{Unconv_1,Unconv_2}.} 
\end{figure}

We can use Uemura's classification \cite{Umera} scheme to place AgSnSe$_2$ in the context of other superconductors based on the ratio of $T_c$/$T_{\mathrm{F}}$. For unconventional superconductors, the ratio generally falls in 0.01 $\leq$ $T_c$/$T_{\mathrm{F}}$ $\leq$ 0.1, however for conventional superconductors, $T_c$/$T_{\mathrm{F}}$ $\leq$ 0.0003. The effective Fermi temperature, $T_{\mathrm{F}}$ of AgSnSe$_2$ is extracted by considering the 3D Fermi surface expression \cite{Tf},
\begin{equation}
k_{B}T_{\mathrm{F}} = \frac{\hbar^{2}}{2}(3\pi^{2})^{2/3}\frac{n^{2/3}}{m^{*}}, 
\label{eqn16:Tf}
\end{equation}
where $n$ is the carrier density, and $m^{*}$ is the effective mass. $m^*$ is calculated from Sommerfeld coefficient via relation, $m^* = {(\hbar k_{\mathrm{F}})^{2}\gamma_{n}/\pi^{2}n k_B^{2}}$, with $\gamma_{n}$ = 84.51 Jm$^{-3}K^{-2}$, $k_{\mathrm{F}}$ is the Fermi vector, and $n$ is taken from \cite{ass2sc}. The estimated value $T_{\mathrm{F}}$ = 16200(610) K places AgSnSe$_2$ well outside the unconventional superconductors band, lying close to conventional superconductors, as shown in \figref{Fig6:UP}. The Uemura classification and no evidence of time-reversal symmetry breaking in AgSnSe$_2$ suggest the conventional superconducting pairing mechanism. 

Doping with valency-skipped elements can enhance or induce superconductivity in low-carrier-density systems such as topological semimetals or semiconductors. This could be a prototype for realizing novel quantum phases called topological superconductivity (TSC), where the topological phase and superconductivity coexist \cite{tsc}. An example is In-doped SnTe, a possible topological superconductor, where the valence-skipped state of In induces superconductivity in a topological crystalline insulator SnTe \cite{ist,ist2}. Moreover, Ag-doped SnSe and other systems, including In-doped GeTe and K-doped BaBiO$_3$, exhibit nontrivial topological band structures \cite{igt, bbo3sc}. Thus, the aforementioned material corresponds to the category of possible topological superconductors and presents AgSnSe$_2$ as a candidate for TSC. However, detailed studies and band structure calculations are required to investigate the origin of multigap superconductivity and to address the effect of the nontrivial band topology with the possible negative-$U$ induced superconductivity in AgSnSe$_2$.
\begin{table}[h!]
\caption{Parameters in the superconducting and normal state of AgSnSe$_2$ }
\begingroup
\setlength{\tabcolsep}{12pt}
\begin{tabular}{c c c} 
\hline\hline
Parameters & Unit & AgSnSe$_2$ \\ [1ex]
\hline
$T_{C}$& K& 4.91(2)\\             
$H_{c1}(0)$& mT& 5.83(6)\\
$H_{c2}^{GL}(0)$& T& 2.14(3)\\
$H_{c2}^{2G}(0)$& T& 2.18(7)\\
$H_{c2}^{P}(0)$& T& 9.13(3)\\
$\xi_{GL}(0)$& nm& 12.2(4)\\
$\lambda_{GL}(0)$ (Mag.)& nm& 309(13)\\
$\lambda(0)$ ($\mu$SR)& nm& 907(35)\\
$k_{GL}(0)$& &25(2)\\
$\gamma_{n}$&  mJ mol$^{-1}$ K$^{-2}$& 4.6(1)\\
$\theta_{D}$& K& 255(3)\\
$\lambda_{e-ph}$ &  & 0.69(1)\\
$\Delta_2/\Delta_1^{SH}$ & & 2.7(3)\\
$\Delta_2/\Delta_1^{\mu SR}$ & & 2.7(5)\\
$T_{\mathrm{F}}$& K& 16200(610)\\
$T_{c}/T_{\mathrm{F}}$& & 0.0003(1)\\
$m^{*}$/$m_{e}$&  & 1.88(2)\\
[1ex]
\hline\hline
\end{tabular}
\endgroup
\end{table}
\section{CONCLUSION}

Detailed measurements of magnetization and specific heat have confirmed the bulk superconductivity of AgSnSe$_2$ at a transition temperature of 4.91(3) K. Microscopic investigations were also carried out using muon spin rotation and relaxation measurements on this valence-skipped superconductor. The results of our TF-$\mu$SR measurements suggest the presence of two isotropic $s + s$ superconducting gaps, which is supported by the temperature variation of specific heat and upper critical field measurements. Furthermore, our ZF-$\mu$SR study confirms the preserved time-reversal symmetry in the superconducting state, suggesting a conventional superconducting state in valence-skipping mediated superconductors, despite the fact that in these compounds, the attractive interaction between electrons is mediated via negative -U due to valence-skipping/valence fluctuations, which is a non-BCS pairing mechanism. However, the role of nontrivial topological states in the superconducting ground state requires further investigation.

\section{Acknowledgments}
A. Kataria acknowledges the funding agency Council of Scientific and Industrial Research (CSIR), Government of India, for providing SRF fellowship (Award No: 09/1020(0172)/2019-EMR-I). R.~P.~S.\ acknowledges Science and Engineering Research Board, Government of India, for the CRG/2019/001028 Core Research Grant. G.M. Luke (McMaster) acknowledges the support of the Natural Sciences and Engineering Research Council (Canada).

\end{document}